\documentclass[pre,10pt,amsmath,amssymb]{revtex4}
\usepackage{graphicx}

\usepackage{amsmath,amsfonts,amssymb,bm}

\begin{document}

\draft
\title{Asymmetric L\'evy flights in nonhomogeneous environments}

\author
{Tomasz Srokowski}

\affiliation{
 Institute of Nuclear Physics, Polish Academy of Sciences, PL -- 31-342
Krak\'ow,
Poland }

\date{\today}

\begin{abstract}

We consider stochastic systems involving general -- non-Gaussian and asymmetric -- stable processes. The random 
quantities, either a stochastic force or a waiting time in a random walk process, explicitly depend on the position. 
A fractional diffusion equation corresponding to a master equation for a jumping process with a variable 
jumping rate is solved in a diffusion limit. It is demonstrated that for some model parameters 
the equation is satisfied in that limit by the stable process with the same asymptotics as the driving noise. 
The Langevin equation containing a multiplicative noise, depending on the position as a power-law, is solved; 
the existing moments are evaluated. The motion appears subdiffusive and the transport depends on the asymmetry 
parameter: it is fastest for the symmetric case. As a special case, the one-sided distribution is discussed. 
\end{abstract} 


\maketitle


\section{Introduction}

Long jumps and divergent fluctuations are observed in various areas: in physics, biology, 
finance and sociology \cite{klag,bou,met}, 
indicating the existence of the power-law tails of the distributions, $\sim|x|^{-1-\alpha}$, where 
$0<\alpha<2$ is a stability index (L\'evy flights). In contrast to the Gaussian distribution, 
the general stable L\'evy distribution can be asymmetric and, in the limiting cases, 
assume a stretched exponential tail. In particular, the distribution may be restricted to a half-axis. 
Specifically the asymmetric processes, where the asymmetry is measured by a 'skewness' parameter $\beta$, 
emerge in many problems and are frequently discussed in the literature. 
A complicated picture of the anomalous diffusion in the reaction-diffusion 
systems is due to asymmetric L\'evy flights and the right-moving fronts accelerate exponentially. 
They develop an algebraic tail, while the left-moving fronts have exponential decaying
tails and move at a constant speed \cite{cas}. A fractional advection-dispersion equation with 
any degree of skewness in several dimensions was analysed in \cite{mee}, whereas 
the problem of the diffusion in the porous media, which display a fractal structure, 
was studied by a stochastic equation driven by the asymmetric L\'evy process in \cite{par}. 
The first passage times for the asymmetric L\'evy flights were evaluated in \cite{koren,magd,dyb2} 
and properties of the stochastic resonance discussed in \cite{dybr,dybr1}. 
A model of a granular material, which takes into account disordered packings of rigid, frictionless disks in two
dimensions under gradually varying stress, predicts a dependence of a strain on the stress direction 
in a form of the asymmetric L\'evy distribution \cite{com}. 
Strongly asymmetric L\'evy flights were observed in cracking of heterogeneous materials \cite{pain,tall}. 
It was demonstrated in the field of finance that prices of the derivatives satisfy a fractional
partial differential equation corresponding to the asymmetric L\'evy processes \cite{cart}. 
In the framework of the jumping processes, a diffusion equation, fractional both in space and time 
and corresponding to a master equation for the random walk with the asymmetric 
L\'evy distribution, was discussed in \cite{goren}; the appropriate algorithms were derived in order to numerically 
simulate the time-evolution. 

Dynamical descriptions of materials containing impurities and defects must take into account random quantities 
which actually do not evolve with time if the time-scale of the impurities 
diffusion is much larger than that of the measured variable (a quenched disorder). 
This introduces a correlation between the successive 
trapping times; the trapping time at a given site is the same for each visit of this site \cite{bou}. 
As a result, the hopping rates are position-dependent: the jump probability function 
in a random walk description has a coupled form and the corresponding Langevin equation possesses 
a multiplicative noise. Usually, the Langevin equation with 
the additive noise is studied and, in such approaches, the medium nonhomogeneity 
is actually reduced to a homogeneous distribution of the noise activation times \cite{sch}. 
Only few studies are devoted to the multiplicative noise. The anomalous diffusion in a composite medium
was studied in terms of a fractional equation with a variable coefficient in \cite{sti}. 
A master equation, describing a thermal 
activation of jumping particles within the folded polymers, also contains a variable diffusion coefficient \cite{gei}. 
A stochastic Lotka-Volterra model has been applied 
to the case when the extreme events exhibit the L\'evy statistics \cite{cog} and a Verhulst 
equation to a population density description \cite{dub}. A recent analysis \cite{bose} of a tumour growth includes 
a coupling between the tumour and an immune cell which leads to a multiplicative noise. 
Traps make the transport slower: anomalous diffusion is a subdiffusion i.e. the variance, if exists, 
rises slower than linearly. The accelerated diffusion, in turn, emerges when 
variance is infinite, due to the long jumps. 

Studies of the Langevin equation with the multiplicative L\'evy noise \cite{sro09,sro10} for the 
symmetric case demonstrate that physical conclusions qualitatively depend on a particular 
interpretation of the stochastic integral. The dynamics in the Stratonovich interpretation may be 
characterised by a finite variance, even in the absence of any potential, and then the solution exhibits fast falling 
power-law tails and a subdiffusive behaviour. On the other hand, the It\^o interpretation predicts 
the same asymptotics the driving noise has. In this paper, we consider a general, asymmetric case 
and demonstrate, in particular, how the asymmetry parameter influences the anomalous transport. 
We begin with the continuous time random walk (CTRW) theory, well-known for 
the L\'evy flights, but usually restricted to homogeneous distributions of the waiting time.

\section{Random walk with a variable jumping rate} 

CTRW is defined in terms of the two distributions: the waiting-time distribution 
$w(t)$ and the jump-length distribution $Q(x)$. Usually, one assumes that they are 
independent stochastic variables (the decoupled version of CTRW) and the resulting 
process appears non-Markovian, except the case of the Poissonian $w(t)$. If 
$w(t)$ has long tails and $Q(x)$ is the Gaussian, the Fokker-Planck equation, which emerges 
from the master equation in the limit of small wave numbers, is fractional in time \cite{met}. 
Then the trapping times hamper the transport and one observes the subdiffusion. When, on the other hand, 
$Q(x)$ obeys the general L\'evy statistics with $\alpha<2$, the variance is infinite. 

We consider the Markovian case and assume, in contrast to the usual approach, that the jumping rate 
depends on the process value: $\nu=\nu(x)$ \cite{kam}. The Markovian property implies a Poissonian form, 
\begin{equation}
\label{poi}
w(t)=w(t|x)=\nu(x){\mbox e}^{-\nu(x)t}, 
\end{equation} 
and by introducing the variable $\nu(x)$ we take into account that the waiting time depends on 
the medium structure. The process is defined by an infinitesimal stationary transition probability, 
  \begin{equation}
  \label{trkp}
p_{tr}(x,\Delta t|x',0)= \{1-\nu(x') \Delta t\}\delta(x-x')+\nu(x') \Delta t Q(x-x').
  \end{equation}
The particle remains at rest for a time sampled from $w(t)$ after which instantaneously jumps 
to a new position and then the process is stepwise constant. 
The first term on the right-hand side of Eq.(\ref{trkp}) is the probability 
that no jump occurred in the time interval $(0,\Delta t)$ and the term $\nu(x') \Delta t$ 
means the probability that one jump occurred.
The master equation derived from Eq.(\ref{trkp}) is the following 
  \begin{equation}
  \label{fpkp}
  \frac{\partial}{\partial t}p(x,t) = -\nu(x)p(x,t) +
  \int Q(x-x')\nu (x') p(x',t) dx'.
  \end{equation} 
The distribution $Q(x)$ represents the L\'evy flights. Trajectories corresponding to that 
kinetics form a self-similar clustering at all scales and exhibit long jumps between clusters; 
such an intermittent behaviour is frequently observed in physical phenomena and modelled by CTRW 
(\cite{dyb2} and references therein). 
$Q(x)=Q_{\alpha,\beta}(x)$ is assumed as a general stable distribution defined by 
the parameters $\alpha$ and $\beta$: $0<\alpha\le 2$ and $|\beta|\le\alpha$ for $0<\alpha<1$ 
and $|\beta|\le 2-\alpha$ for $1<\alpha<2$. The case $\alpha=1$ is special and will not be 
considered; we also neglect parameters related to translation and scaling 
of the distribution. The characteristic function has the form 
\begin{equation}
\label{fq}
\widetilde Q_{\alpha,\beta}(k)=\exp[-|k|^\alpha\exp\left(i\frac{\pi\beta}{2}\hbox{sign}(k)\right)]
\end{equation}
and the density follows from the inverse Fourier transform, 
\begin{equation}
\label{ffq}
Q_{\alpha,\beta}(x)=\frac{1}{\pi}\hbox{Re}\int_0^\infty \widetilde Q_{\alpha,\beta}(k){\mbox e}^{-ikx}dk. 
\end{equation}
Eq.(\ref{fpkp}) is known as the Kolmogorov-Feller equation for a pure discontinuous Markovian process \cite{sai}. 
It can be derived from the Langevin equation as a generalisation of the Kolmogorov equation for the Markovian 
non-Gaussian processes \cite{dub2,dub3}. 

A description of CTRW in terms of a fractional, both in space and time, differential equation 
is possible in the diffusion limit, i.e. for small arguments of the Fourier-Laplace transform \cite{met}. 
In that limit, the characteristic function reads,  
\begin{equation}
\label{dl}
\widetilde Q_{\alpha,\beta}(k)\approx 1-|k|^\alpha\exp\left(i\frac{\pi\beta}{2}\hbox{sign}(k)\right)+{\mbox O}(k^2). 
\end{equation}
Transforming Eq.(\ref{fpkp}) and applying Eq.(\ref{dl}) yields the equation, 
\begin{equation}
\label{fracek}
\frac{\partial\widetilde p(k,t)}{\partial t}=-|k|^\alpha\exp\left(i\frac{\pi\beta}{2}\hbox{sign}(k)\right)
{\cal F}[\nu(x)p(x,t)], 
\end{equation} 
that leads, after applying Eq.(\ref{ffq}) and for $\nu(x)$=const, to a fractional differential 
equation \cite{yan}. The nonhomogeneity of the medium is reflected by the $x-$dependence 
of the jumping rate: the sojourn time of the particle in a trap 
depends on the position and then the diffusion coefficient in Eq.(\ref{fracek}) is variable. We assume 
\begin{equation}
\label{nu}
\nu(x)=K|x|^{-\theta}, 
\end{equation}
where $K$ has been introduced for dimensional reasons; in the following we take $K=1$. 
In particular, when $\theta>0$ the average jumping rate is large near the origin whereas the average 
waiting time becomes large at large distances. The power-law form of $\nu(x)$ is natural for problems 
exhibiting self-similarity, which often happens for disordered materials; it has been applied e.g. to study 
diffusion on fractals \cite{osh,met3} and turbulent two-particle diffusion \cite{fuj}. 
To solve Eq.(\ref{fracek}) we assume $\alpha>1$ but a generalisation to the case $\alpha<1$ is straightforward. 

There is no general method to exactly solve the fractional equation with a variable diffusion coefficient.  
However, since the equation itself was derived by applying the condition (\ref{dl}), 
we may restrict our considerations to the limit $|k|\to 0$ without introducing any additional 
idealisation. The solution is not unique since one can construct, in principle, a family of solutions 
the characteristic functions of which differ at orders higher than $|k|^\alpha$. Can this family include 
the stable distributions? We will demonstrate that it is indeed the case but only in a limited range of 
the model parameters. The stable distribution always can be expressed 
in a form of the Fox $H$-function with well-determined coefficients \cite{schn,mat,sri}. 
Therefore, the variability of the diffusion coefficient 
may manifest itself in the solution only as a time-dependent scaling factor and the required solution of Eq.(\ref{fracek}) 
with the initial condition $p(x,0)=\delta(x)$ must have the form 
\begin{eqnarray} 
\label{solp}
p(x,t)=\frac{\epsilon}{\sigma(t)^\epsilon}H_{2,2}^{1,1}\left[\frac{x}{\sigma(t)^\epsilon}\left|\begin{array}{l}
(1-\epsilon,\epsilon),(1-\gamma,\gamma)\\
\\
(0,1),(1-\gamma,\gamma)
\end{array}\right.\right],
\end{eqnarray} 
where $\epsilon=1/\alpha$, $\gamma=(\alpha-\beta)/2\alpha$ and the function $\sigma(t)$ is to be determined. 
The derivation, presented in Appendix A, shows that (\ref{solp}) satisfies Eq.(\ref{fracek}) to the lowest order in $|k|$ 
and yields a power-law time-dependence 
\begin{equation}
\label{sodt}
\sigma(t)=\left[A(\alpha+\theta)t\right]^{\alpha/(\alpha+\theta)},
\end{equation}
where
$$
A=\frac{2}{\pi\alpha^2}\Gamma(\theta/\alpha)\Gamma(1-\theta)
\sin\left(\frac{\pi\theta}{2}\right)\cos\left(\frac{\beta\theta}{2\alpha}\pi\right)
$$
and $-\alpha<\theta<1$. The latter inequality specifies the conditions under which it is possible to express 
the solution of Eq.(\ref{fracek}), valid in the diffusion limit, in the form of the stable process. 
The asymptotic form of the distribution, $\sim|x|^{-1-\alpha}$, follows from the expansion of the 
$H$-function in Eq.(\ref{solp}); it is the same as that of the driving noise. 
The solution (\ref{solp}) satisfies the following scaling relations, 
\begin{equation}
\label{scre}
t\to\lambda t, ~~x\to\lambda^\kappa x ~~ \hbox{and} ~~ p(x,t)\to \lambda^{\kappa}p(\lambda^\kappa x,\lambda t), 
\end{equation}
where $\kappa=1/(\alpha+\theta)$ and $\lambda>0$ is an arbitrary scaling parameter. On the other hand, 
the scaling relations (\ref{scre}) can be directly inferred from the fractional equation (\ref{fracek}). 

Eq.(\ref{fracek}) can be solved also for $\theta\ge1$ and for that purpose the $H$-function coefficients 
must be modified similarly to the symmetric case \cite{sro06,pha}. However, then we would leave a domain 
of the stable distributions. 

Asymptotic shape of the solution indicates that all moments of the order $\delta\ge\alpha$ diverge and 
the transport properties may be quantified by fractional moments. 
The existing fractional moments, corresponding to the solution (\ref{solp}), are given by the Mellin transform 
from the $H$-function, 
\begin{equation}
\label{mom}
\langle|x|^\delta\rangle=\frac{1}{\alpha}\sigma(t)^{\delta/\alpha}[\chi_+(-\delta-1)+\chi_-(-\delta-1)]=
-\frac{2\sigma(t)^{\delta/\alpha}}{\pi\alpha}\Gamma(-\delta/\alpha)\Gamma(1+\delta)
\sin\left(\frac{\pi\delta}{2}\right)\cos\left(\frac{\beta\delta}{2\alpha}\pi\right),
\end{equation}
where $\chi_\pm(-s)$ denotes the Mellin transform corresponding to $\pm\beta$, 
and $\langle x\rangle=0$; therefore, $\langle|x|^\delta\rangle\sim t^{\delta/(\alpha+\theta)}$. 
The proportionality coefficient, presented in Fig.1 as a function of $\beta$ for some values of $\theta$, 
rises with $\theta$ and a maximum always corresponds to the symmetric case. 
For $\theta=0$, $\langle|x|^\delta\rangle(\beta)$ has a cosine shape. 

The divergence of the variance may be unphysical if one considers the motion of a massive particle 
though this property does not violate physical principles for such problems as the diffusion in energy space 
in spectroscopy or for the diffusion on a polymer chain in the chemical space \cite{met}. To get rid 
of the difficulty of divergent moments one introduces the L\'evy walk \cite{met} which relates the jump length 
to the velocity. It is possible to generalise the above model by introducing a dependence of 
the velocity variance on time; then one obtains a strong anomalous diffusion and the scaling 
relations different from those for the ordinary L\'evy walk \cite{and}. 
One the other hand, one can argue that every system is finite and introduce a truncation 
of the distribution in a form of a fast-falling tail. The truncated distribution agrees with the L\'evy 
distribution up to an arbitrarily large value of the argument and has the finite variance. The problem 
of the truncated L\'evy flights for the multiplicative processes (in the symmetric case) 
was discussed in Ref.\cite{pha}. Variance is always finite for $\alpha=2$ and then all kinds of diffusion emerge. 
In particular, for $\theta<0$, we observe the enhanced diffusion, 
$\langle|x|^\delta\rangle\sim t^{\delta/(2+\theta)}$, which represents a strong anomalous 
diffusion in a sense of Ref.\cite{cast}. 
\begin{center}
\begin{figure}
\includegraphics[width=12cm]{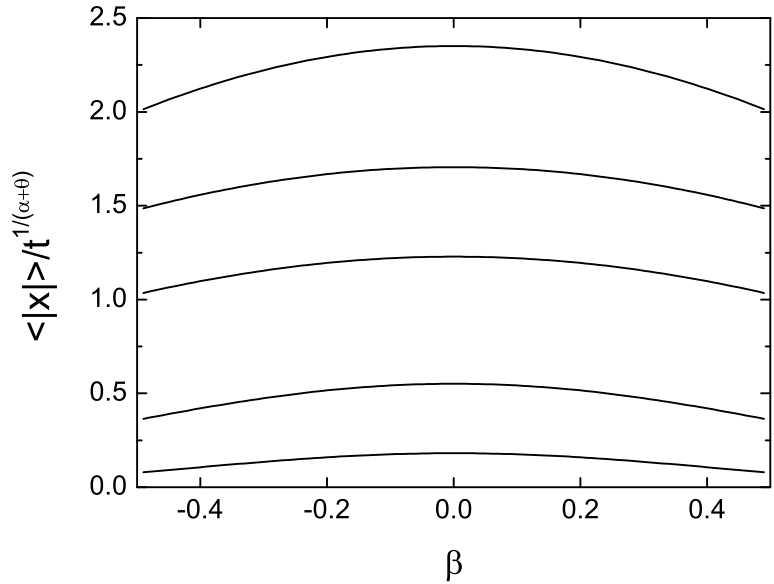}
\caption{The fractional moments as a function of the asymmetry parameter calculated 
from Eq.(\ref{mom}) for $\delta=1$, $\alpha=1.5$ and the following values of $\theta$: -1.2, -1, -0.5, 0 and 0.5 
(from bottom to top).}
\end{figure}
\end{center} 

\section{Langevin equation}

The stochastic dynamics driven by a multiplicative, algebraic random force and a linear deterministic force 
is governed by the Langevin equation, 
\begin{equation}
\label{la}
dx=-\lambda xdt+K|x|^{-\theta/\alpha}d\eta(t), 
\end{equation} 
where we assume that the increments $d\eta(t)$ are distributed according to (\ref{ffq}), $\lambda\ge0$ 
and the constant $K=1$ cm$^{\theta/\alpha}$ will be dropped in the following. Since $\eta$ 
represents the white noise, Eq.(\ref{la}) requires a clarification of how a stochastic integral 
is to be interpreted \cite{gar}. In the It\^o interpretation, which is frequently used just due to its simplicity, 
the noise term is evaluated before the noise acts and applies 
when the noise consists of clearly separated pulses; this is e.g. the case of CTRW. Moreover, 
it has been demonstrated both for the Gaussian \cite{kup} and the general 
L\'evy case \cite{sro12} that this interpretation is suited for problems with a large inertia. 
On the other hand, the Stratonovich interpretation, which takes into account a middle point 
between the subsequent noise activations, applies if a system has finite correlations and 
the white noise is only an approximation. For the Gaussian case, the difference between the above interpretations 
resolves itself to a drift term in the corresponding Fokker-Planck equation. 

The dynamics involving the multiplicative noise and an arbitrary potential can be expressed 
in the It\^o interpretation by a fractional equation with a variable diffusion 
coefficient \cite{sch} which in our case takes the form \cite{yan,uwa}, 
\begin{equation}
\label{frace}
\frac{\partial p}{\partial t}=-\lambda\frac{\partial}{\partial x}(xp)-(-\Delta)^{\alpha/2}(|x|^{-\theta}p)
+\tan(\pi\beta/2)\frac{\partial}{\partial x}(-\Delta)^{(\alpha-1)/2}(|x|^{-\theta}p), 
\end{equation}
where 
\begin{equation}
\label{pfr}
(-\Delta)^{\alpha/2}f(x)={\cal F}^{-1}(|k|^\alpha\widetilde f(k)). 
\end{equation}
Eq.(\ref{frace}) involves, beside a usual fractional diffusion term -- 
present in the Fokker-Planck equation for the symmetric case -- 
a contribution to the convection due to existence of the preferred direction \cite{yan}. 
The equation which determines the characteristic function is a generalisation of Eq.(\ref{fracek}): 
\begin{equation}
\label{lak}
\frac{\partial\widetilde p(k,t)}{\partial t}=-\lambda k\frac{\partial}{\partial k}
{\widetilde p}(k,t)-|k|^\alpha\exp\left(i\frac{\pi\beta}{2}\hbox{sign}(k)\right)
{\cal F}[|x|^{-\theta}p(x,t)].
\end{equation} 
We look for a solution in the diffusion limit of small $|k|$ and apply 
a procedure similar to that in Sec.II. The solution with the initial condition $p(x,0)=\delta(x)$ 
is given by Eq.(\ref{solp}) and $\sigma(t)$ satisfies the equation 
\begin{equation}
\label{rsi}
\dot\sigma(t)=-\alpha\lambda\sigma(t)+\alpha A\sigma(t)^{-\theta/\alpha} 
\end{equation}
which has the solution
\begin{equation}
\label{sodto}
\sigma(t)=\left[\frac{A}{\lambda}(1-{\mbox e}^{-\lambda(\alpha+\theta)t}\right]^{1/c_\theta}
\end{equation} 
where $c_\theta=1+\theta/\alpha$. 
$p(x,t)$ converges with time to a stationary state and the tails $\sim|x|^{-1-\alpha}$ ($|\beta|<2-\alpha$) 
make the variance divergent for any $t$. Numerical values of $p(x,t)$ can be obtained by expanding 
the $H$-function near $x=0$ and $|x|=\infty$, by means of Eq.(\ref{A.5}). The case $|\beta|=2-\alpha$ 
has a different asymptotics which is presented in Appendix B. 
On the other hand, the density distributions can be calculated from a direct numerical integration of Eq.(\ref{la}). 
Fig.2 presents examples of such distributions, close to the stationary states, where the driving noise 
was sampled according to a standard procedure \cite{wer}. The tails indicate a power-law shape 
except the case $\beta=2-\alpha$ when the left tail falls faster than exponentially. 

We have already mentioned that an important property of the above solution 
is that all moments of the order $\ge\alpha$ diverge. 
The existence of the infinite variance was reported for some physical problems, e.g. for the rain and clouds 
fields \cite{sch1}; according to that study, the experimental radar rain reflectivities 
indicate the divergence of all moments higher than the value 1.06. However, the infinite 
variance is often unphysical and its presence in such problems as the advecting field for porous media 
and atmospheric turbulence has been questioned \cite{sch}. Usually, one observes the power-law distributions 
with the tails falling faster than for the stable L\'evy distributions. This is the case for the financial 
market \cite{sta,ple,gab} and the minority game \cite{ren}; fast falling power-law tails result 
from a multifractal analysis of the extreme events \cite{muz} and emerge when one introduces a power-law truncation 
to the distribution \cite{sok,pha}. We shall demonstrate that Eq.(\ref{la}) predicts such fat tails, 
without introducing any truncation, but in a different interpretation of the stochastic integral. 

\begin{center}
\begin{figure}
\includegraphics[width=12cm]{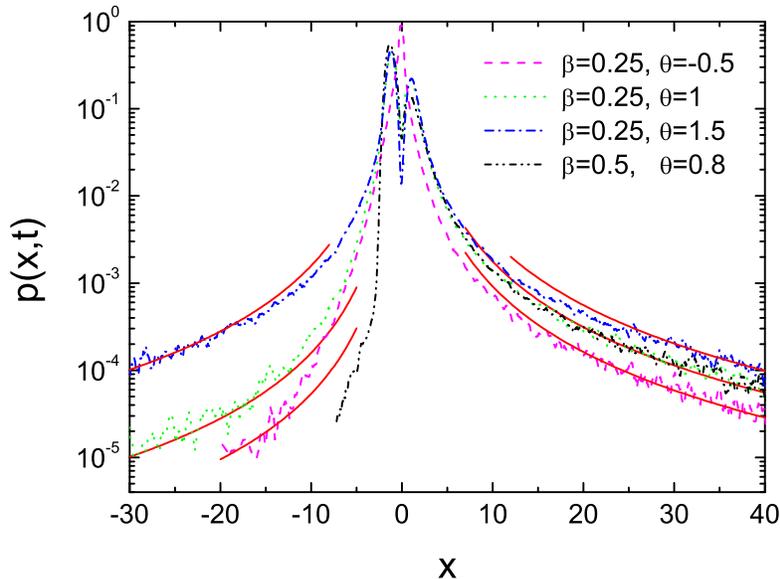}
\caption{The density distributions calculated by integration of Eq.(\ref{la}) 
in the It\^o interpretation for $\lambda=1$, $\alpha=1.5$ 
and $t=5$. The solid lines mark the dependence $|x|^{-1-\alpha}$. For each curve, $10^7$ trajectories was evaluated.}
\end{figure}
\end{center} 

The physical importance of the Stratonovich interpretation, in which the random driving is evaluated at a middle point 
between its consecutive activations, stems from the fact that it corresponds to a white-noise 
limit of the coloured noises. Then the usual change of the variable leads to the Langevin equation with the additive noise 
-- for one-dimensional systems and for the Gaussian noise \cite{gar}. 
If $\alpha<2$, one can formally define the white noise $\eta$ as a limit of a coloured noise: construct a coloured-noise 
process, change the variable and finally take the white-noise limit. This procedure can be easily accomplished 
for the generalised Ornstein-Uhlenbeck process, 
\begin{equation}
\label{kolo}
d\eta_c(t)=-\gamma_n\eta_c(t)dt+\gamma_ndL(t),  
\end{equation}
where $dL(t)$ has the stable L\'evy distribution \cite{sro12}; then 
$d\eta(t)$ is given by a limit of the vanishing relaxation time, $d\eta(t)=\lim_{\gamma_n\to\infty}d\eta_c(t)$. 
The numerical analysis for the symmetric noise demonstrates \cite{sro09,sro10} that results obtained by means 
of the variable transformation agree with those for the white noise in the Stratonovich interpretation. 
Then we solve the equation 
\begin{equation}
\label{laso}
\dot y=-\lambda c_\theta y+\eta(t), 
\end{equation} 
obtained from Eq.(\ref{la}) by the transformation  
\begin{equation}
\label{yodx}
y(x)=\frac{1}{c_\theta}|x|^{c_\theta}\hbox{sign}(x), 
\end{equation} 
assuming $\alpha+\theta>0$. Eq.(\ref{laso}) is easy to solve \cite{jes} and applying the identity 
$p(x,t)=p(y(x),t)dy/dx$ yields the final solution. Since the higher and lower domain 
of $\alpha$ are qualitatively different, it is expedient to consider them separately. 

\subsection{The case $\alpha>1$}

The solution of Eq.(\ref{laso}) for this case can be expressed in the same form as Eq.(\ref{solp}) \cite{schn}. 
After the variable transformation, the solution of Eq.(\ref{la}) for $x>0$ reads 
\begin{eqnarray} 
\label{solsx1}
p(x,t)=\frac{c_\theta}{\alpha x}H_{2,2}^{1,1}\left[\frac{x^{c_\theta}}
{c_\theta\sigma_s(t)^{1/\alpha}}
\left|\begin{array}{l}
(1,1/\alpha),(1,\gamma)\\
\\
(1,1),(1,\gamma)
\end{array}\right.\right], 
\end{eqnarray} 
where 
\begin{equation}
\label{sodtos}
\sigma_s(t)=\frac{1-{\mbox e}^{-\lambda(\alpha+\theta)t}}{\lambda(\alpha+\theta)}, 
\end{equation} 
and one should change $\beta\to-\beta$ to get the solution for $x<0$. The numerical values 
of $p(x,t)$ for small $|x|$ follow from the $H$-function expansion, Eq.(\ref{A.5}). The derivation 
yields the series: 
\begin{equation}
\label{szm1}
p(x,t)=\frac{c_\theta}{\pi}\sum_{n=1}^\infty c_\theta^{-n}\sigma_s(t)^{-n/\alpha}\Gamma(n/\alpha)\sin(\pi n\gamma)
\frac{(-1)^{n-1}}{(n-1)!}x^{c_\theta n-1}. 
\end{equation}
Similarly, after transforming the argument of the $H$-function $x\to x^{-1}$, we get the asymptotic expansion, 
\begin{equation}
\label{szd1}
p(x,t)=\frac{c_\theta}{\pi}\sum_{n=1}^\infty c_\theta^{n\alpha}\sigma_s(t)^n\Gamma(n\alpha)\sin(\pi n\alpha\gamma)
\frac{(-1)^{n-1}}{(n-1)!}x^{-1-(\alpha+\theta)n}. 
\end{equation}
Therefore, the asymptotic form of the distribution, $\sim|x|^{-1-\alpha-\theta}$, differs from the 
It\^o version: the slope depends on $\theta$ and may be arbitrarily large. 
The above results are valid for $\beta\ne|\alpha-2|$; otherwise the distribution falls 
faster than exponentially (see Appendix B). Though all the integer moments higher than the fist of this extremely asymmetric 
process are still infinite, the existence of the right-sided Laplace transform for $\beta=\alpha-2$
makes it useful for the applications e.g. in finance, where it is known as the FMLS model. 
A particular feature of the FMLS process is that it only exhibits downwards jumps, 
while upwards movements have continuous paths \cite{cart}. 
\begin{center}
\begin{figure}
\includegraphics[width=12cm]{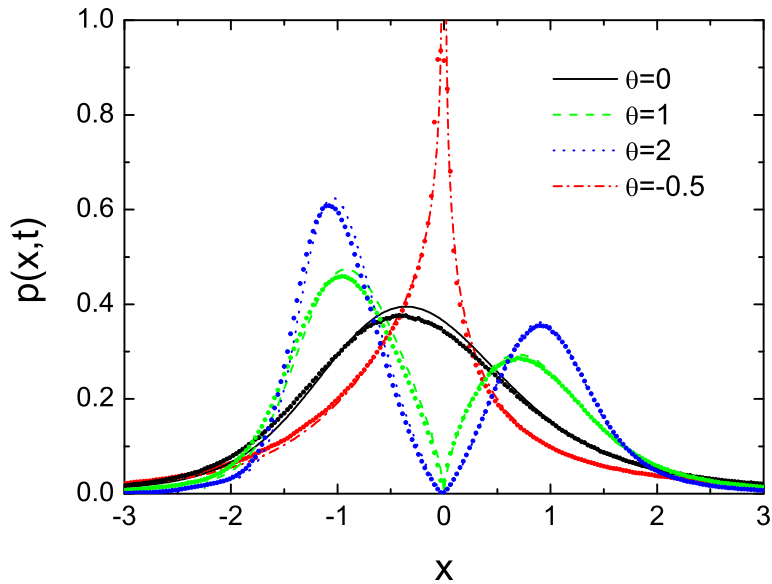}
\caption{The density distributions calculated from Eq.(\ref{szm1}) (lines) and 
by integration of Eq.(\ref{laso}) (points) for $\lambda=1$, $\alpha=1.5$, $\beta=0.25$  
and $t=5$.}
\end{figure}
\end{center} 

Fig.3 presents a comparison of the analytical distributions, Eq.(\ref{szm1}), 
with those obtained from the numerical simulations for some values of $\theta$; 
a relatively large time, sufficient to reach the stationary state, was chosen. For a positive 
$\theta$ the density vanishes at $x=0$ and the peak is split, in contrast to 
$\theta<0$ when $p(0,t)$ is infinite. 
\begin{center}
\begin{figure}
\includegraphics[width=12cm]{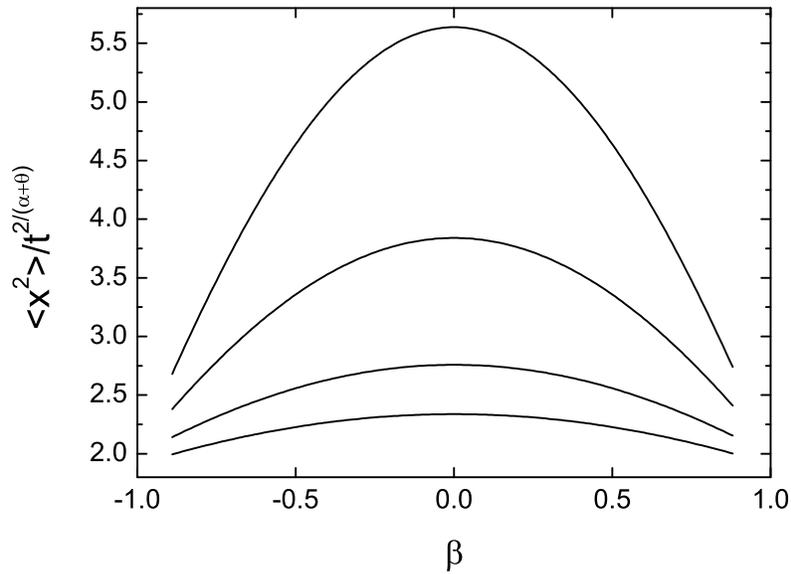}
\caption{The variance as a function of the asymmetry parameter calculated 
from Eq.(\ref{var}) for $\alpha=1.1$ and the following values of $\theta$: 1.5, 2, 3 and 4 
(from top to bottom).}
\end{figure}
\end{center}

The dependence of the distribution slope on $\theta$ makes this parameter responsible for the existence 
of the moments, in contrast to the It\^o case, and that property 
implies important consequences for the diffusion ($\lambda=0$). 
It may be accelerated, as it was for the It\^o interpretation, but if $\theta$ is chosen sufficiently large, 
the moment of an arbitrary high order exists, namely $n-$th moment exists if $\theta>n-\alpha$. Let 
us evaluate the variance assuming $\alpha+\theta>2$. Using the Mellin transform from the $H$-function yields 
\begin{equation}
\label{var}
\langle x^2\rangle(t)=\frac{1}{\alpha}c_\theta^{2/c_\theta}t^{2/(\alpha+\theta)}[\chi_+(-2/c_\theta)+\chi_-(-2/c_\theta)]=
-\frac{2}{\pi\alpha}c_\theta^{2/c_\theta}\Gamma(1+2/c_\theta)\Gamma(-2/(\alpha+\theta))
\sin(\pi/c_\theta)\cos\left(\frac{\beta\pi}{\alpha+\theta}\right)t^{2/(\alpha+\theta)}. 
\end{equation}
Therefore, the variance, if exists, always rises with time slower than linearly and we observe the subdiffusion. 
Eq.(\ref{var}) is illustrated with Fig.4. The coefficient $\langle x^2\rangle/t^{2/(\alpha+\theta)}$ 
has a cosine shape as a function of $\beta$ and its value strongly depends on $\theta$ near $\beta=0$. 
On the other hand, the dependence on $\theta$ is relatively weak for the strongly asymmetric cases. 

The linear growth with time of the variance (the normal diffusion) is expected on time scales 
larger than a microscopic time scale and is a consequence of the 
central limit theorem. If, on the other hand, correlations decay slowly, that theorem does not apply and 
the anomalous transport emerges. Subdiffusion in CTRW results from a particle trapping which effect increases 
with time since the waiting time distribution has long tails \cite{met}; the medium in that theory is regarded as 
homogeneous. The anomalous transport predicted by Langevin equation with the multiplicative noise 
has a different origin: it results from the noise intensity which decreases with the distance. 
Not only the multiplicative noise is able to increase the distribution slope and make the variance finite; 
this also happens when one introduces a nonlinear deterministic force into the Langevin equation \cite{che,sro10}. 
However, then a stationary state exists and the subdiffusion does not occur. 
The finite moments have been found in the Verhulst model for the population density in which the random force 
is multiplicative and given by the one-sided L\'evy distribution \cite{dub1}. Moreover, they emerge 
due to the trapping inside a potential when the dynamics is driven by short overdamped Josephson junctions 
and distributions of the noise signals have long tails \cite{aug}. 

The anomalous diffusion is a generic property of the complex systems and emerges in many fields \cite{met}. 
In those systems nonhomogeneity effects are important and the central limit theorem does not apply. 
It is the case, in particular, for transport in the media with a quenched disorder \cite{bou}, 
as well as in the biological systems: in a cytoplasm and cellular membranes, where macromolecules 
are densely packed and exhibit heterogeneous structures; the {\it in vitro} experiments reveal 
the subdiffusion in those systems (for a recent review see \cite{hof}). 

\subsection{The case $\alpha<1$}

Very long jumps, i.e. possessing the infinite first moment, are also observed in realistic 
physical systems. For example, the molecular dynamics 
calculations in the framework of a granular material model \cite{com}, in which rigid spheres 
are densely and randomly packed under gradually varying stress, yield the L\'evy-distributed large strain 
increments with a value of $\alpha$ in the 0.4-0.6 range. 
A numerical handling of the dynamics driven by a noise with $\alpha<1$ is difficult. 
Moreover, those processes can be non-ergodic: 
it has been proved \cite{bark} that a weak non-ergodic behaviour 
emerges in the CTRW when the the waiting-time distribution is given by a one-sided L\'evy distribution. 

Now, the $H$-function representation of the stable distribution is different than 
for the case $\alpha>1$ \cite{schn} and it leads to the following solution of Eq.(\ref{la}), 
\begin{eqnarray} 
\label{solsx2}
p(x,t)=\frac{c_\theta}{\alpha x}H_{2,2}^{1,1}\left[\frac{x^{c_\theta}}
{c_\theta\sigma_s(t)^{1/\alpha}}
\left|\begin{array}{l}
~~~~(2,1),(2\gamma,\gamma)\\
\\
(2/\alpha,1/\alpha),(2\gamma,\gamma)
\end{array}\right.\right], 
\end{eqnarray} 
where $\gamma\ne$0, 1. The series expansions for small and large arguments are the same as for $\alpha>1$, 
they are given by Eq.(\ref{szm1}) and (\ref{szd1}), respectively. Also 
an expansion for the intermediate values can be derived (see \cite{ben} for the symmetric case). 
\begin{center}
\begin{figure}
\includegraphics[width=12cm]{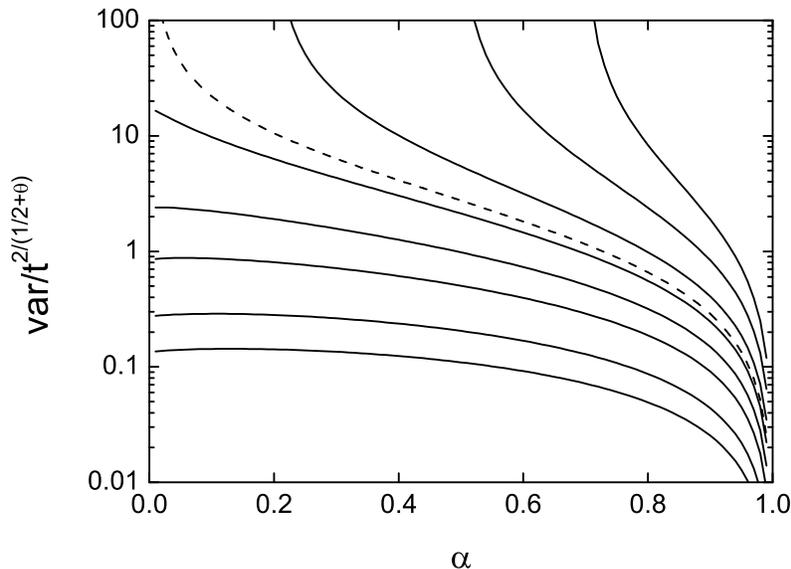}
\caption{The variance for the one-sided case $\beta=-\alpha$ as a function of $\alpha$ calculated 
from Eq.(\ref{v1s}) for the following values of $\theta$: 1.3, 1.5, 1.8, 2, 2.1, 2.5, 3, 4 and 5 
(from top to bottom). The limiting case $\theta=2$ is marked by the dashed line.}
\end{figure}
\end{center}

The cases $\beta=\alpha$ and $\beta=-\alpha$ represent one-sided distributions, restricted to $x<0$ and $x>0$, 
respectively. Those maximally asymmetric L\'evy flights are useful to describe 
multifractal processes with $\alpha<2$ \cite{sch2}, applied e.g. as a model 
of the atmospheric phenomena \cite{sch1}. The one-sided processes were discussed from the point of view 
of the first passage time and first passage leapover problems in \cite{eli,koren}. 
For $\beta=-\alpha$, the terms on the main diagonal in Eq.(\ref{solsx2}) are identical and can be eliminated; 
the reduction formula of the $H$-function yields 
\begin{eqnarray} 
\label{sols1s}
p(x,t)=\frac{c_\theta}{\alpha x}H_{1,1}^{1,0}\left[\frac{x^{c_\theta}}
{c_\theta\sigma_s^{1/\alpha}}
\left|\begin{array}{l}
~~~(2,1)\\
\\
(2/\alpha,1/\alpha)
\end{array}\right.\right]. 
\end{eqnarray} 
Similarly to the two-sided case, the $n-$th moment exists if $\alpha+\theta>n$ and now the mean value does not 
vanish. The variance $\hbox{var}(t)=\langle(x-\langle x\rangle)^2\rangle$ for $\lambda=0$, 
given by the Mellin transform from Eq.(\ref{sols1s}), rises with time slower than linearly:  
\begin{equation}
\label{v1s}
\hbox{var}(t)=\frac{1}{\alpha}c_\theta^{2/c_\theta}\left[\frac{\Gamma(-2/(\alpha+\theta))}{\Gamma(-2/c_\theta)}-\frac{1}{\alpha}
\frac{\Gamma^2(-1/(\alpha+\theta))}{\Gamma^2(-1/c_\theta)}\right]t^{2/(\alpha+\theta)}. 
\end{equation}
The expression (\ref{v1s}) is illustrated in Fig.5 for some values of $\theta$ as a function of $\alpha$. 
The variance rapidly decreases with $\theta$ and the dependence on $\alpha$ is weak for large $\theta$, 
except a vicinity of $\alpha=1$. If $\theta>2$, $\hbox{var}(t)<\infty$ for any $\alpha$. 
\begin{center}
\begin{figure}
\includegraphics[width=12cm]{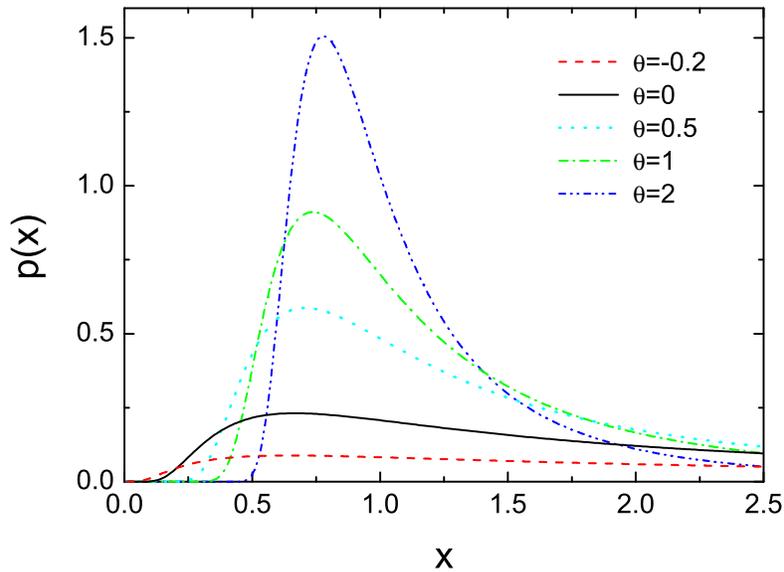}
\caption{The stationary distributions calculated from Eq.(\ref{sa12}) 
for $\alpha=1/2$ and $\lambda=1$.} 
\end{figure}
\end{center} 

The $H$-function becomes an elementary function for $\alpha=1/2$ and $\beta=-1/2$ (the L\'evy-Smirnov distribution): 
\begin{equation}
\label{sa12}
p(x,t)=\frac{\sigma_s(t)}{2\sqrt{\pi}}(1+2\theta)^{3/2}x^{-3/2-\theta}\exp\left(-\frac{1}{4}(1+2\theta)\sigma_s(t)^2 x^{-1-2\theta}\right). 
\end{equation}
The distributions corresponding to the stationary states are presented in Fig.6. They exhibit a maximum at 
$x_m=[\lambda^2(3/2+\theta)]^{-1/(1+2\theta)}$ 
which rises with $\theta$ and shifts towards $x=1$. Then $\lim_{\theta\to\infty}p(x,t)=\delta(x-1)$ for any $t>0$ 
and we observe an instantaneous jump $x=0\to x=1$. The curves corresponding to small $\theta$, in particular negative, 
disclose a uniform pattern with long tails. The diffusive case, $\lambda=0$, also is characterised by a strongly localised 
density for large $\theta$ and the peak moves with time to infinity, $x_m\sim t^{1/(1/2+\theta)}$. 

\section{Summary and conclusions} 

We discussed stochastic processes driven by the asymmetric stable distributions and 
the medium nonhomogeneity was taken into account 
by introducing a multiplicative noise. Process defined in that way may no longer be stable. 
However, in the diffusion limit of small wave numbers, when the master equation for a jumping 
process becomes the fractional Fokker-Planck equation, one can approximate them by the stable processes. 
We have considered a coupled version of CTRW where the jumping rate depends on the position as a 
power-law function, $|x|^{-\theta}$, and demonstrated that such an approximation 
is valid but only in a limited range of the parameters: $-\alpha<\theta<1$. 
The resulting density has the same form as the jump-size distribution with the asymptotic slope independent 
of $\beta$ and $\theta$; only its time-dependence 
is affected by the variable jumping rate. This property is in contrast to the Gaussian case, $\alpha=2$, 
characterised by a stretched exponential asymptotics. All fractional moments of the order $\delta\ge\alpha$ 
are infinite for any $\beta$ and $\theta$, they are largest for the symmetric case. 

The Fokker-Planck equation for the coupled CTRW is fractional and has a variable diffusion 
coefficient. In the Langevin formulation of CTRW, the stochastic force is multiplicative 
and the equation requires the It\^o interpretation. On the other hand, the solutions 
of the Langevin equation obtained by a transformation of the process variable -- 
a procedure which corresponds to the Stratonovich interpretation --  have different properties: 
they can possess finite moments. 
If the variance is finite, i.e. if $\alpha+\theta>2$, it rises with time slower than linearly: 
$\langle x^2\rangle(t)={\cal D}t^{2/(\alpha+\theta)}$. We observe a subdiffusion as a result 
of diminishing of the noise intensity with the distance. The coefficient ${\cal D}$ is largest for the symmetric case
and then strongly depends on $\theta$. On the other hand, the very asymmetric cases exhibit a moderate 
growth of ${\cal D}$ with $\theta$. The above conclusions can be generalised to other forms 
of the multiplicative noise in Eq.(\ref{la}) if they have a sufficiently large slope. 

The extremely asymmetric case for $\alpha<1$ corresponds to a one-sided distribution which also 
can possess the finite variance. If $\theta$ is 
sufficiently large, ${\cal D}$ appears almost constant as a function of $\alpha$ in a wide range 
of this parameter but the variance rapidly falls to zero for $\alpha\to1$. The one-sided case has been 
illustrated with the L\'evy-Smirnov density. The role of the multiplicative noise is clearly visible 
for this simple process: it makes the distribution wide and uniform when $\theta$ is negative, 
whereas for large values of $\theta$ the distribution shrinks to the delta function. 

It is not a priori clear which of the two solutions of Eq.(\ref{la}), (\ref{solp}) or (\ref{solsx1}), 
applies to a concrete physical system; the main difference between them consists in a different slope 
of the tails. Some experimental work would be helpful. Shape of the probability distributions 
can be determined experimentally and such studies, applied to the heterogeneous systems, could reveal 
effects of the variable diffusion coefficient. From that point of view, the analysis of fractures 
of the disordered materials and measuring the crack propagation is promising since in those systems 
complex processes proceed on a broad range of scales and exhibit self-similar properties. A recent 
study \cite{tall} demonstrates that the experimental distribution of the local velocities 
of the crack front is characterised by a power-law tail, $v^{2.7}$, and the global velocity distribution 
converges by upscaling to the asymmetric stable distribution for scales larger than the spatial correlation 
length. At smaller scales, in turn, only the tail agrees with the stable distribution 
which conclusion may indicate, in view of our results, the It\^o interpretation of Eq.(\ref{la}) and 
presence of the strong nonhomogeneity ($|\theta|$ relatively large). The analysis presented in this paper 
shows namely that $p(x,t)$ can converge to the stable distribution only if $\nu(x)$ is sufficiently smooth; 
otherwise, $p(x,t)$ still possesses a fat tail corresponding to the stable distribution
but behaviour near $x=0$ may depend on $\theta$ \cite{sro06}. 
A characteristic feature of systems governed by Eq.(\ref{la}) ($\lambda=0$) is a specific 
time-dependence of the fractional moments, similar for both interpretations, which can both decrease 
and increase the transport speed, compared to the homogeneous case. This property appears robust 
in respect to the nonhomogeneity parameter $\theta$ and has been observed not only for the stable 
solutions (\ref{solp}) \cite{sro06}. It has been argued in Ref. \cite{tall} that the experimentally 
estimated variance of the global crack front velocity assumes a finite value since the system 
is actually finite. It would be interesting to verify whether that quantity, measured for a small resolution 
to make the self-similar structure apparent, obey Eq.(\ref{mom}).

\section*{APPENDIX A}

\setcounter{equation}{0}
\renewcommand{\theequation}{A\arabic{equation}} 

We will show that the function (\ref{solp}) satisfies Eq.(\ref{fracek}) to the lowest order in $|k|$ 
and evaluate $\sigma(t)$. First, the Fourier transform from both $p(x,t)$ and $p_\theta(x,t)\equiv|x|^{-\theta}p(x,t)$ is needed. 
Since for any stable density $f_\beta(-x)=f_{-\beta}(x)$, we only consider $x\ge 0$. Then we have, 
\begin{equation}
\label{A.1}
\widetilde f_\beta(k)=\int_0^\infty[(f_\beta(x)+f_{-\beta}(x))\cos(kx)+i(f_\beta(x)+f_{-\beta}(x))\sin(kx)]dx 
\end{equation}
and $\int_0^\infty f(x)\sin(kx)dx=-\frac{\partial}{\partial k}\int_0^\infty x^{-1}f(x)\cos(kx)dx$. 
The cosine transform from the $H$-function is given by the general formula: 
  \begin{eqnarray} 
\label{A.2}
\int_0^\infty H_{p,q}^{m,n}\left[x\left|\begin{array}{c}
(a_p,A_p)\\
\\
(b_q,B_q)
\end{array}\right.\right]\cos(kx)dx= 
\frac{\pi}{k}H_{q+1,p+2}^{n+1,m}\left[k\left|\begin{array}{l}
(1-b_q,B_q),(1,1/2)\\
\\
(1,1),(1-a_p,A_p),(1,1/2)
\end{array}\right.\right].
  \end{eqnarray} 
Moreover, the multiplication rule, 
  \begin{eqnarray} 
\label{A.3}
x^\sigma H_{p,q}^{m,n}\left[x\left|\begin{array}{c}
(a_p,A_p)\\
\\
(b_q,B_q)
\end{array}\right.\right]= 
H_{p,q}^{m,n}\left[x\left|\begin{array}{c}
(a_p+\sigma A_p,A_p)\\
\\
(b_q+\sigma B_q,B_q)
\end{array}\right.\right], 
  \end{eqnarray} 
yields 
  \begin{eqnarray} 
\label{A.4}
p_\theta(x,t)=\sigma(t)^{-\epsilon(1+\theta)}H_{2,2}^{1,1}\left[\frac{x}{\sigma(t)^\epsilon}\left|\begin{array}{l}
(1-\epsilon-\theta\epsilon,\epsilon),(1-\gamma_\pm-\theta\gamma_\pm,\gamma_\pm)\\
\\
(-\theta,1),(1-\gamma_\pm-\theta\gamma_\pm,\gamma_\pm)
\end{array}\right.\right], 
\end{eqnarray}
where $\gamma_\pm=(\alpha\mp\beta)/2\alpha$. Then we take the Fourier transform from Eq.(\ref{A.4}); 
the existence of $\widetilde p_\theta(k,t)$ requires that the singularity at $x=0$ must not be essential 
which, in turn, implies the condition $\theta<1$. 

Next, we expand all the functions in Eq.(\ref{A.1}), for $p$ and $p_\theta$, in powers of $|k|$ by applying the general formula, 
  \begin{eqnarray} 
\label{A.5}
H_{p,q}^{m,n}\left[z\left|\begin{array}{c}
(a_p,A_p)\\
\\
(b_q,B_q)
\end{array}\right.\right]=\sum_{h=1}^m\sum_{\nu=0}^\infty\frac{\prod_{j=1,j\ne
h}^m\Gamma(b_j-B_j\frac{b_h+\nu}{B_h})\prod_{j=1}^n\Gamma(1-a_j+A_j\frac{b_h+\nu
}{B_h})}{\prod_{j=m+1}^q\Gamma(1-b_j+B_j\frac{b_h+\nu}{B_h})\prod_{j=n+1}^p
\Gamma(a_j-A_j\frac{b_h+\nu}{B_h})}\frac{(-1)^\nu z^{(b_h+\nu)/B_h}}{\nu!B_h},
  \end{eqnarray} 
which is valid if $|\beta|<2-\alpha$ and $\sum B_i>\sum A_i$; the latter condition is satisfied 
for the case $\alpha>1$. Evaluating the first two terms in (\ref{A.1}), corresponding to 
the l.h.s. of Eq.(\ref{fracek}), yields 
\begin{equation}
\label{A.6}
\hbox{Re }\widetilde p(k,t)=1+\pi (h_\alpha^++h_\alpha^-)\sigma^{-1}|k|^\alpha+O(k^2),
\end{equation} 
where the coefficient 
\begin{equation}
\label{A.7}
h_\alpha^\pm=-\frac{\alpha}{2\pi}\cos(\pi\alpha/2)\sin(\pi\alpha\gamma_\pm)/\sin(\pi\alpha)
\end{equation}
corresponds to the term 
$h=2$ and $\nu=1$ in Eq.(\ref{A.5}). A similar calculation for the imaginary part yields
\begin{equation}
\label{A.8} 
\hbox{Im }\widetilde p(k,t)=-\alpha\sigma\sin(\pi\beta/2)|k|^\alpha+O(k^2). 
\end{equation}

To evaluate the r.h.s of Eq.(\ref{fracek}) up to the required order, it is sufficient to determine the term $k^0$ 
which is real. The simplest method makes use of the Mellin transform, namely: 
\begin{equation}
\label{A.9}
\widetilde p_\theta(k=0,t)=\int_{-\infty}^\infty p_\theta(x,t)dx=\sigma^{-\theta/\alpha}[\chi_+(-1)+\chi_-(-1)]=
\frac{2\sigma^{-\theta/\alpha}}{\pi}\Gamma(1-\theta)\Gamma(\theta/\alpha)\sin(\pi\theta/2)\cos\left(\frac{\beta\theta}{2\alpha}\pi\right),
\end{equation}
where $\chi_\pm(-s)$ denotes the Mellin transform corresponding to $\pm\beta$. 
Since, asymptotically, $p_\theta\sim |x|^{-1-\alpha-\theta}$, the convergence of the integral in Eq.(\ref{A.9}) imposes  
the condition $\alpha+\theta>0$. Introducing the above results to Eq.(\ref{fracek}) produces two identical equations 
for the real and imaginary parts which determine the function $\sigma(t)$, 
\begin{equation}
\label{A.10}
\dot\sigma(t)=\frac{2}{\pi\alpha}\Gamma(\theta/\alpha)\Gamma(1-\theta)
\sin\left(\frac{\pi\theta}{2}\right)\cos\left(\frac{\beta\theta}{2\alpha}\pi\right)\sigma(t)^{-\theta/\alpha},
\end{equation}
and its solution with the initial condition $\sigma(0)=0$ is given by Eq.(\ref{sodt}). 

\section*{APPENDIX B}

\setcounter{equation}{0}
\renewcommand{\theequation}{B\arabic{equation}} 

For $\beta=\alpha-2$ the $H$-function in Eq.(\ref{solp}) can be reduced to a lower order and the density expressed as 
\begin{eqnarray} 
\label{B.1}
p(x,t)=\frac{\epsilon}{\sigma(t)^\epsilon}H_{1,1}^{1,0}\left[\frac{x}{\sigma(t)^\epsilon}\left|\begin{array}{l}
(1-\epsilon,\epsilon)\\
\\
(0,1)
\end{array}\right.\right].
\end{eqnarray} 
Since $n=0$, the power-law asymptotics is no longer valid and instead an exponential behaviour emerges \cite{bra}. 
In the case of Eq.(\ref{B.1}), the Fox function has the following form for the large arguments, 
\begin{equation}
\label{B.2}
H(z)=c_1z^\lambda{\mbox e}^{-c_2z^{c_3}}, 
\end{equation}
where $c_1=[2\pi(\alpha-1)\alpha^{1/(\alpha-1)}]^{-1/2}$, $c_2=(\alpha-1)\alpha^{-c_3}$, $c_3=\alpha/(\alpha-1)$ and 
$\lambda=(2-\alpha)/[2(\alpha-1)]$; the distribution falls faster than exponentially. 

A similar reduction applies to the case $\alpha<1$ and the result, similar to Eq.(\ref{B.2}), 
represents an expansion for small arguments \cite{schn}. If $\alpha$ is a rational number, the distribution 
for the one-sided cases can be expressed by the generalised hypergeometric functions and then relatively easy 
evaluated \cite{gor}.


\begin{thebibliography}{99}
\bibitem{uwa}
Eq.(\ref{frace}) is not unique because the definition of the fractional operator itself 
is not unique \cite{yan}.
\end{thebibliography}
\end{document}